\documentclass[a4paper,twocolumn]{esapub} % European paper
\usepackage[sectionbib]{natbib}
\usepackage{graphicx}
\usepackage{xspace}
\usepackage{times}
\usepackage[dvips]{color}

\newcommand{\e}[1]{\ifmmode{%
\mathchoice{^{\mbox{\scriptsize #1}}}{^{\mbox{\scriptsize #1}}}%
{^{\mbox{\tiny #1}}}{^{\mbox{\tiny #1}}}}%
\else{$^{\mbox{\scriptsize #1}}$}\fi}
\renewcommand{\i}[1]{\ifmmode{%
\mathchoice{_{\mbox{\scriptsize #1}}}{_{\mbox{\scriptsize #1}}}%
{_{\mbox{\tiny #1}}}{_{\mbox{\tiny #1}}}}%
\else{$_{\mbox{\scriptsize #1}}$}\fi}
\newcommand{\ei}[2]{\ifmmode{%
\mathchoice{^{\mbox{\scriptsize #1}}_{\mbox{\scriptsize #2}}}%
{^{\mbox{\scriptsize #1}}_{\mbox{\scriptsize #2}}}%
{^{\mbox{\tiny #1}}_{\mbox{\tiny #2}}}%
{^{\mbox{\tiny #1}}_{\mbox{\tiny #2}}}}%
\else{$^{\mbox{\scriptsize #1}}_{\mbox{\scriptsize #2}}$}\fi}

\newcommand{\ev}{e\kern -0.11em V\xspace}
\newcommand{\kev}{ke\kern -0.09em V\xspace}
\newcommand{\keV}{\kev}

\newcommand{\ca}{\mbox{$\sim$}}
\newcommand{\asm}{\mbox{\textsl{ASM}}\xspace}
\newcommand{\ginga}{\mbox{\textsl{Ginga}}\xspace}
\newcommand{\hexe}{\mbox{\textsl{Mir-HEXE}}\xspace}
\newcommand{\intg}{\mbox{\textsl{INTEGRAL}}\xspace}
\newcommand{\isgri}{\mbox{\textsl{ISGRI}}\xspace}
\newcommand{\jemx}{\mbox{\textsl{JEM-X}}\xspace}
\newcommand{\omc}{\mbox{\textsl{OMC}}\xspace}
\newcommand{\sax}{\mbox{\textsl{BeppoSAX}}\xspace}
\newcommand{\spi}{\mbox{\textsl{SPI}}\xspace}
\newcommand{\xte}{\mbox{\textsl{RXTE}}\xspace}
\newcommand{\hd}{\mbox{HD\,77581}\xspace}
\newcommand{\vela}{\mbox{Vela~X-1}\xspace}
\newcommand{\err}[2]{\ensuremath{^{+#1}_{-#2}}\xspace}
\newcommand{\redchi}{\ensuremath{\chi\ei{2}{red}}\xspace}
\newcommand{\xspec}{{\sl XSPEC}\xspace}

\newcommand{\url}[1]{\parbox[t]{0.4\textwidth}{\texttt{#1}}}
\newcommand{\Rsun}{\ensuremath{\mbox{R}_\odot}\xspace}
\newcommand{\Msun}{\ensuremath{\mbox{M}_\odot}\xspace}

\newlength{\tabwidth}
\begin{document}

\title{INTEGRAL broadband spectroscopy of Vela X-1}

\author[1,3]{P.~Kretschmar}
\author[2]{R.~Staubert}
\author[2,3]{I.~Kreykenbohm}
\author[3]{M.~Chernyakova}
\author[1]{A.~v.~Kienlin}
\author[4]{S.~Larsson}
\author[1,3]{K.~Pottschmidt}
\author[2,5]{J. Wilms}
\author[6]{L.~Sidoli}
\author[7]{A.~Santangelo}
\author[7]{A.~Segreto}
\author[8]{D.~Attie}
\author[8]{P.~Sizun}
\author[8]{S.~Schanne}

\affil[1]{Max-Planck-Institut f\"ur Extraterrestrische Physik, 
                 87548 Garching, Germany}
\affil[2]{Institut f\"ur Astronomie und Astrophysik -- Astronomie,
                 Univ. of T\"ubingen, 72076 T\"ubingen, Germany}
\affil[3]{INTEGRAL Science Data Center, 1290 Versoix, Switzerland}
\affil[4]{Department of Astronomy, Stockholm University, SE-10691 Stockholm, Sweden}
\affil[5]{Department of Physics, University of Warwick, 
                 CV4~7AL Warwick, UK}
\affil[6]{Istituto di Astrofisica Spaziale e Fisica Cosmica -- CNR, 
          20133 Milano, Italy}
\affil[7]{Istituto di Astrofisica Spaziale e Fisica Cosmica -- CNR, 
          90146 Palermo, Italy}
\affil[8]{Service d'Astrophysique, CEA--Saclay,           91191 Gif-sur-\kern -0.1em Yvette, France}

\maketitle

%\vspace*{-5mm}

\begin{abstract}
The wind-accreting X-ray binary pulsar and cyclotron line source \vela has
been observed extensively during \intg Core Program observations of the Vela
region in June-July and November-December 2003. In the latter set of
observations the source showed intense flaring -- see also \cite{Staubert:IWS5},
these proceedings.

We present early results on time averaged and time resolved spectra, of both 
epochs of observations. A cyclotron line feature at \ca 53\,\kev is clearly
detected in the \intg spectra and its broad shape is resolved in \spi spectra.
The remaining issues in the cali\-bration of the instruments do not allow to 
resolve the question of the disputed line feature at ~20-25\,\kev.

During the first main flare the average luminosity increases by a factor of
\ca 10, but the spectral shape remains very similar, except for a moderate
softening.

\end{abstract}
%\normalsize

\section{Introduction}\label{Intro}
\vela (4U\,0900$-$40) is an eclipsing high mass X-ray binary with an 
orbital period of 8.96437\,days \citep{Barziv:2001} at a distance of
$\sim$2.0\,kpc \citep{Nagase:89} consisting of the B0.5Ib supergiant 
\hd and a neutron star. The optical companion has a mass of 
$\sim$23\,\Msun and a radius of $\sim$30\,\Rsun while the neutron star
mass is estimated to be \ca 1.8\Msun \citep{Barziv:2001}.

Due to the small separation of the binary system (orbital radius: 
1.7\,R$_\star$), the neutron star is deeply embedded in the intense stellar wind
\citep[4$\times$10\e{$-$5}\,\Msun/yr;][]{Nagase:86} of
\hd. X-ray line spectra measurements \citep{Sako:99} show 
that this wind is inhomogeneous with dense clumps embedded in a much
thinner, highly ionized component.
The typical X-ray luminosity of \vela is \ca 4$\times$10\e{36}\,erg/s,
but both sudden flux reductions to less than 10\,\% of its normal value
\citep{Inoue:84,Lapshov:92,Kreykenbohm:99} and flaring activity 
\citep{Kendziorra:90,HaberlWhite:90,Kreykenbohm:99}
have been observed in the past.

The neutron star has a spin period of $\sim$283\,s
\citep{McClintock:76}.  Both spin period and period
derivative have changed erratically since the first measurement as is
expected for a wind accreting system. The last measurements with the
Burst and Transient Source Experiment\footnote{See~\hfill
  \url{http://www.batse.msfc.nasa.gov/batse/\linebreak[4]pulsar/data/sources/velax1.html}}
resulted in a period of \ca283.5\,s.

The broadband X-ray spectrum of \vela has the typical shape of accreting
pulsar spectra with a power law continuum at lower and an exponential cutoff
at higher energies. This is further modified by strongly varying absorption
which depends on the orbital phase of the neutron star
\citep{Kreykenbohm:99,HaberlWhite:90}, an iron fluorescence line at
6.4\,\kev, and occasionally an iron edge at 7.27\,\kev
\citep{Nagase:86}. A cyclotron resonant scattering feature (CRSF) at
\ca 55\,\kev was first reported from observations with \hexe
\citep{Kendziorra:92}. \citet{Makishima:92} and \citet{Choi:96}
reported an absorption feature at \ca25\,\kev to 32\,\kev from \ginga.
This lower energy feature has been disputed by \sax observations
\citep{Orlandini:98} but supported by phase resolved analysis of
\hexe \citep{Kretschmar:97} and \xte data \citep{Kreykenbohm:2002}.
%\newpage

\begin{figure*}
\hspace*{5mm}
\includegraphics[width=0.40\textwidth]{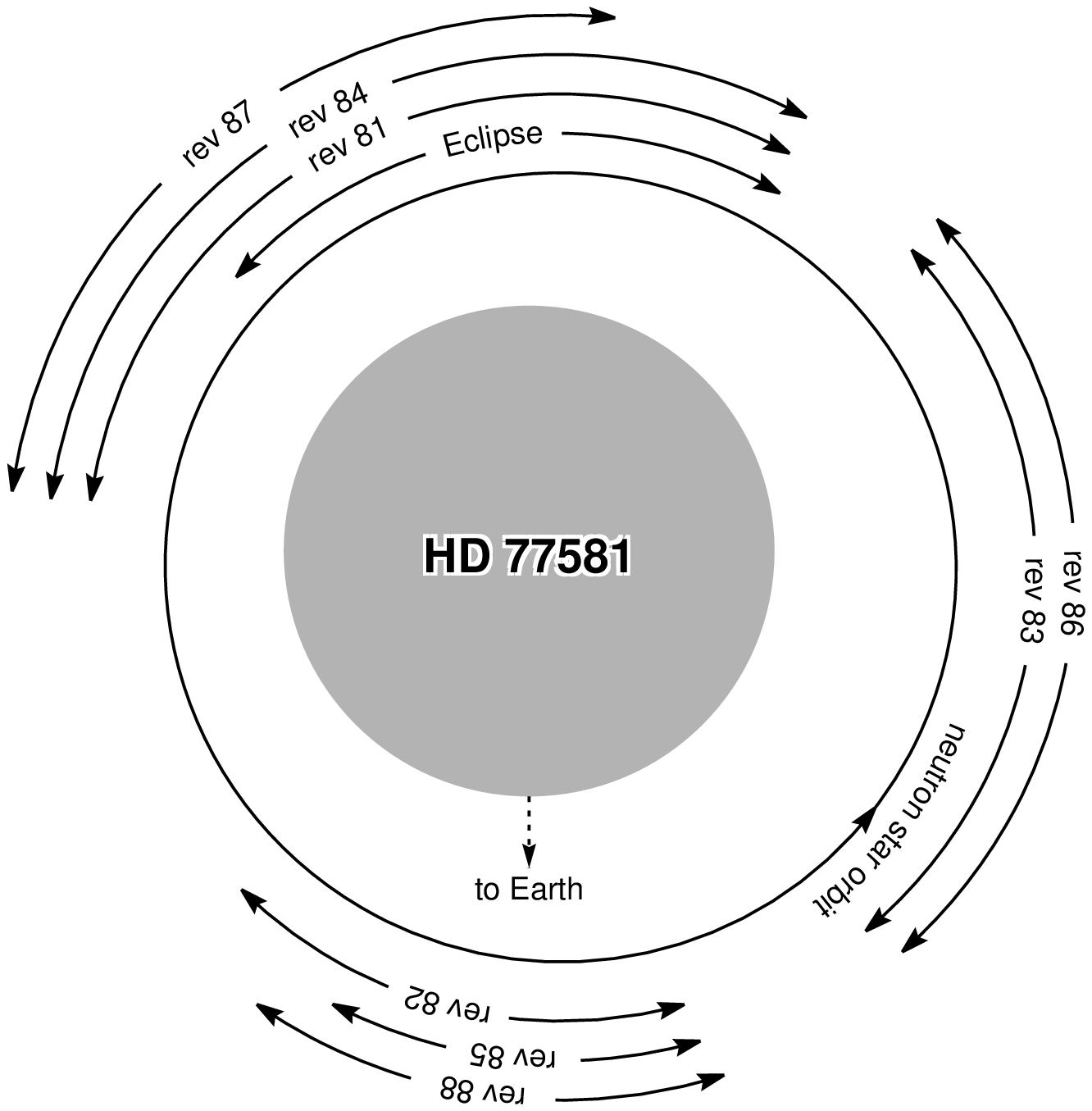}
\hfill 
\includegraphics[width=0.397\textwidth]{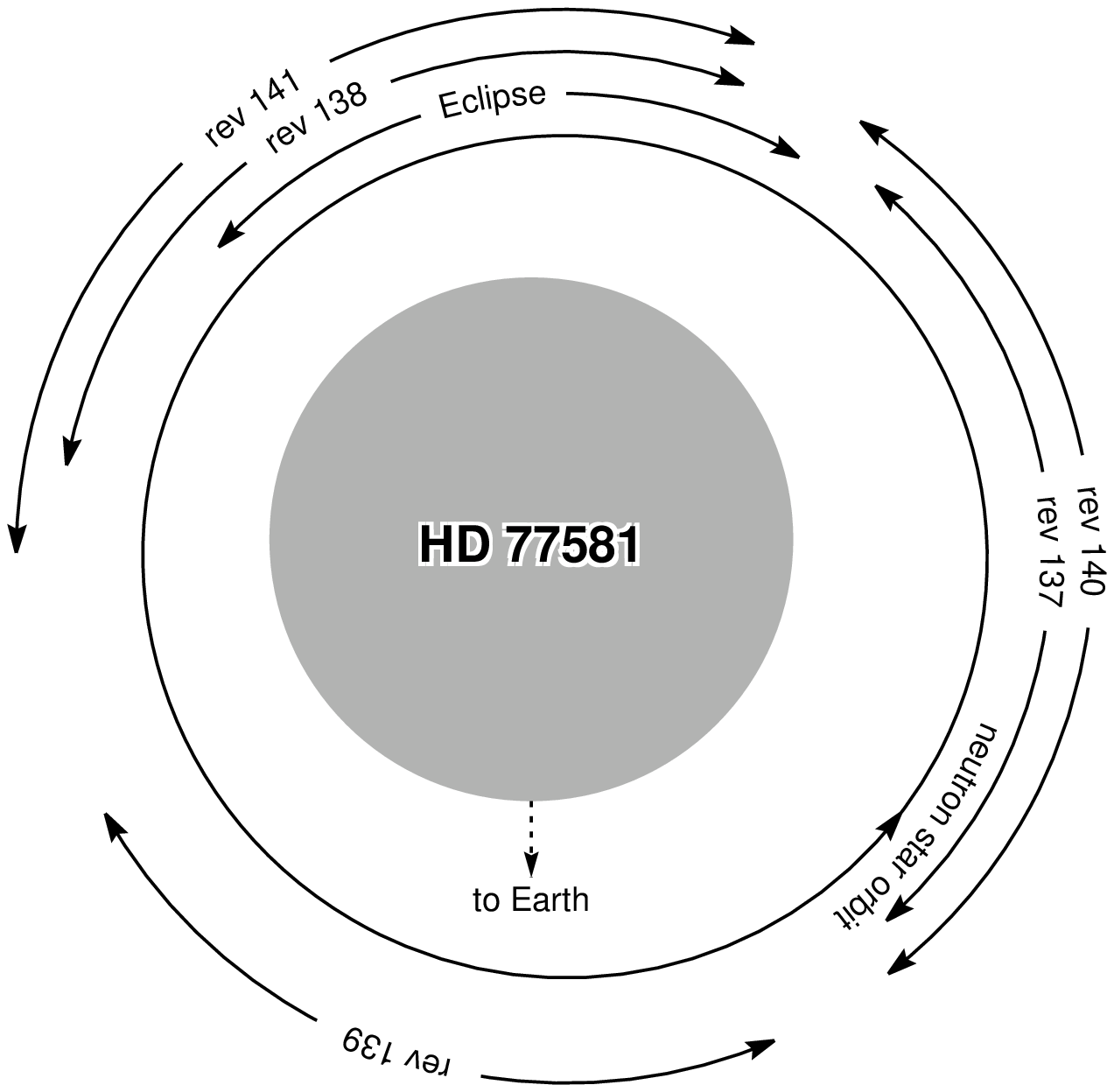}
\hspace*{5mm}

\caption{Sketch of the Vela X-1 System during the observations in summer
         and in winter 2003. The source position along its orbit 
         during the \intg observations is indicated for each 
         \intg revolution.\label{Fig:System}}
\end{figure*}

\begin{figure}
\centerline{\includegraphics[width=0.31\textwidth]{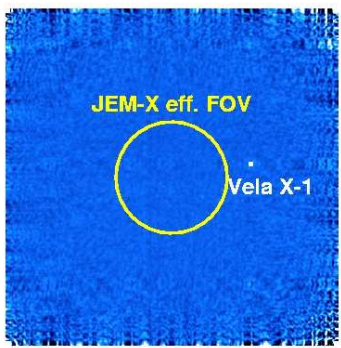}}
\caption{\isgri image of pointing 013700420010 during which the first
         major \vela flare reached its peak. The source was outside of 
         the field of view of the monitor instruments.\label{Fig:FOV}}

\end{figure}

\section{Observations and Data Reduction}\label{Obs}
As part of the \intg Core Program the Vela region has been observed
twice in 2003 for an extended time. The first set of observations
took place mid June to early July (revolutions 81 to 88) the
second end of November to mid December (revolutions 137 to 141).

As the observations were not targeted at any specific source in 
order to support different scientific goals, the observation strategy
was not optimized for \vela. In practice this meant that several
revolutions were scheduled with \vela in eclipse (see Fig.~\ref{Fig:System})
and that for a significant fraction of the time the source was not
within the field-of-view of the two monitor instruments \jemx and \omc.

The two observation epochs, summer and winter, saw a very different behaviour
of \vela. While in summer the source was mostly calm, the winter observations
showed several large flares \citep{Krivonos:ATel211,Staubert:IWS5}, possibly
the largest ever observed. These flares and the general source behaviour
during the observations are described in more detail in \citet{Staubert:IWS5}.
Unfortunately the observation strategy explained above means that we have
no coverage for the flares with the monitor instruments. Also the
\xte-\asm did not sample this source region during the flares.

For the summer observations we exctracted long term average spectra for all
three high energy instruments, excluding times of eclipse and pointings with
other problems like, e.g., radiation belt entries.  While these spectra only
describe an average source state and the exact selection varies from
instrument to instrument, the large total exposure allows to determine a
statistically significant SPI spectrum beyond 50\,\kev despite the steeply
falling source spectrum and degradation of the data by a large solar flare
which occured in revolutions 82/83. The total integration times for the
combined spectra are 340.2 and 333.2~ksec for \jemx and \isgri, respectively,
where concurrent observations were chosen and 775.6~ksec for \spi.

\begin{figure}
\centerline{\includegraphics[angle=-90,width=0.40\textwidth]{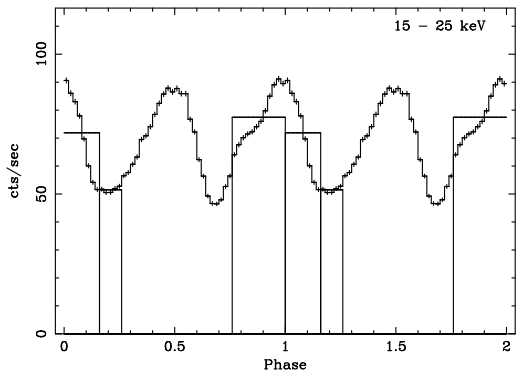}}
\caption{The phase intervals chosen for the pulse phase resolved spectroscopy.\label{Fig:PPS}}

\end{figure}

For the winter observations we have concentrated on the spectra before
and during the first big flare. In this case we have generated pulse
phase resolved spectra covering the rise, the peak and falling flank
and the following off-pulse region connected with the so-called ``main
pulse'' as defined, e.g., by \citet{Kreykenbohm:2002}. The used bins
are indicated in Fig.~\ref{Fig:PPS}.

The data reduction was done with the \textsl{OSA\,3} release of the scientific
analysis software and the corresponding cali\-bration files. Due to the 
remaining problems in the \isgri response matrix with ``snake-like''
residuals, we assumed systematic uncertainties of 10\% for the extracted
\isgri spectra. For \jemx a known problem exists in reconstructing fluxes
below \ca 10\,keV, in our case this leads to artificially enhanced intrinsic
absorption in the, typically strongly absorbed, source.

%\pagebreak[4]
Since no direct support for pulse phase resolved analysis is yet available in
the analysis software, we constructed specific GTI information for the
different phase bins. These GTIs are based on the barycenter and binary
corrected pulse period determination \citep{Staubert:IWS5} from which then
times in the instrumental time frame were derived.

\section{Results}\label{Res}
%\subsection{Summer observations}\label{Res:summer}

The combined spectra of \jemx and \spi 
for the summer observations
can be well fitted at lower
energies by a power law model with exponential cutoff (\xspec model
\texttt{CUTOFFPL}),
% with a so-called Fermi-Dirac cutoff 
%\citep{Tanaka:86},
if one includes free relative normalization between the two spectra,
to allow for \mbox{(inter-)}cali\-bration uncertainties.
A powerlaw with a Fermi-Dirac cutoff \citep{Tanaka:86}
%as used, e.g., by \citet{Kreykenbohm:99} 
works equally well and gives very similar results, but the onset of
the cutoff is not well determined, thus we use the simpler model
here. 

\begin{figure}

%INCLUDE SPECTRUM + RATIO1 + RATIO2 PLOT
\hspace*{-4mm}\includegraphics[angle=-90,width=0.50\textwidth]{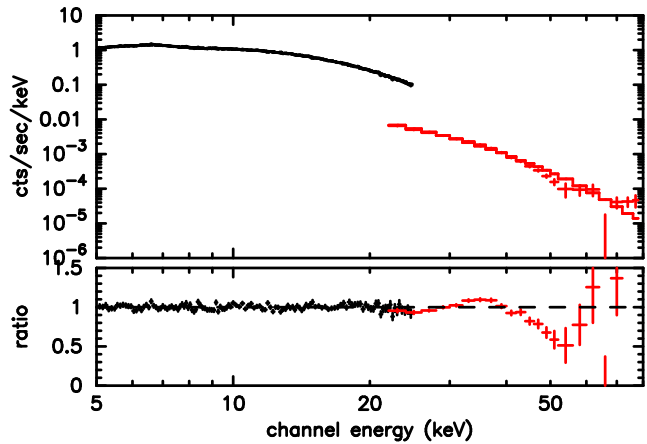}

\caption{\label{Fig:Summer_JS}Combined \jemx and \spi spectra averaged
         over the good pointings during the summer observations.}
\end{figure}

Above 50\,\kev there is an evident structure in the fit residuals for
\spi, see Fig.~\ref{Fig:Summer_JS}. This structure can be fitted by
including a cyclotron scattering feature at $E$$\approx$54\,\keV with
$\sigma$$\approx$7\,\kev. Since
the SPI energy resolution at these energies is \ca 1.6\,\kev 
\citep{Attie:2003}, the line shape can actually be resolved
as far as statistics permit it. The line parameters do not depend
siginificantly on the exact continuum -- simple exponential cutoff
or Fermi-Dirac cutoff -- used.

\begin{figure}

\hspace*{-4mm}\includegraphics[angle=-90,width=0.50\textwidth]{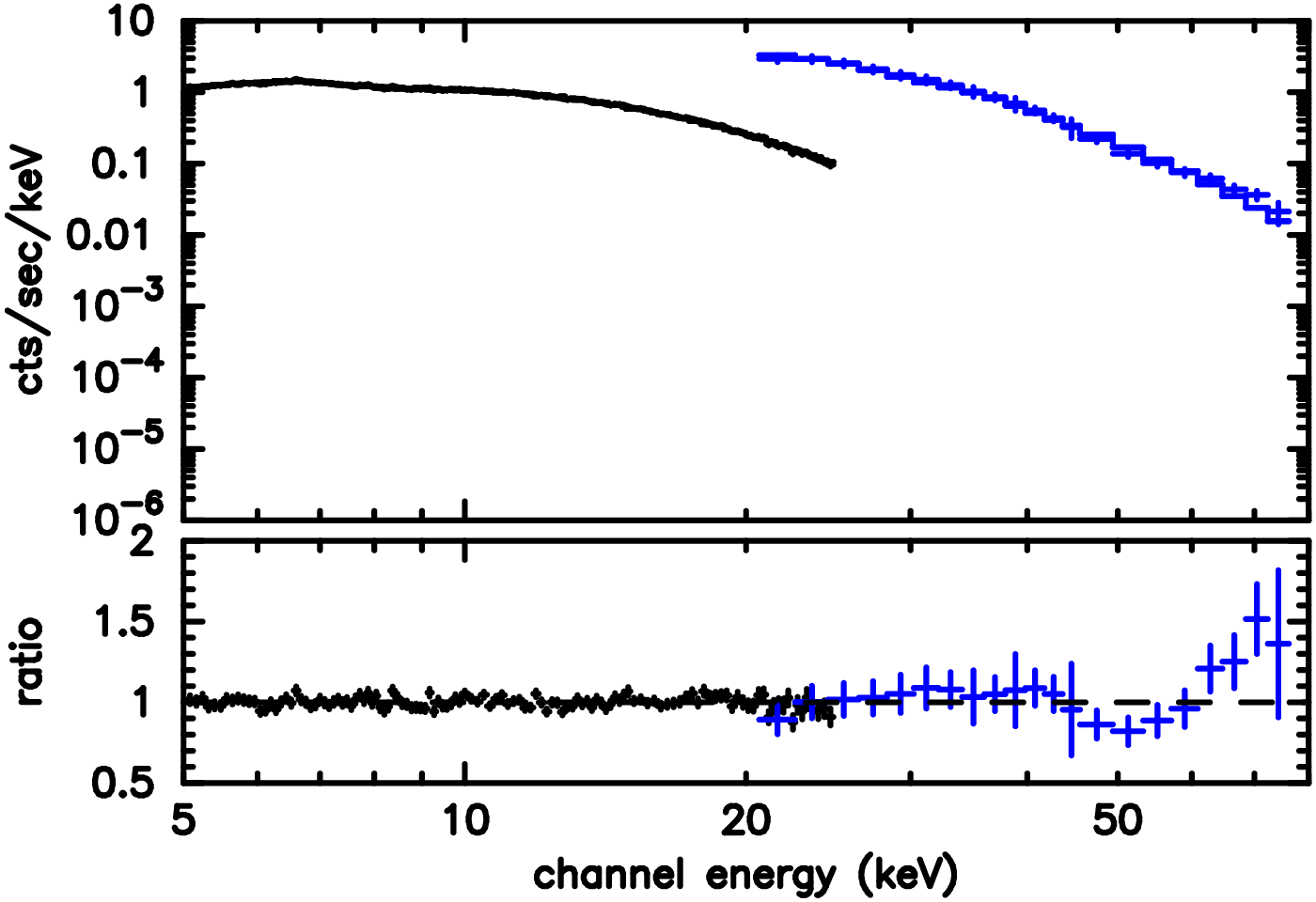}
%INCLUDE SPECTRUM + RATIO1 + RATIO2 PLOT

\caption{\label{Fig:Summer_JI}Combined \jemx and \isgri spectra averaged
         over the good pointings during the summer observations.}
\end{figure}

Combining \jemx and \isgri spectra, the missing flux above \ca 50\,\kev
is again clearly visible in the fit residuals -- 
see Fig.~\ref{Fig:Summer_JI} -- and can be fitted with a line
at {$E$}{$\approx$}52\,\keV. 
Combining spectra from all three high energy instruments does not
allow a good fit due to calibration differences between \isgri and
\spi even though the fit parameters are mostly compatible (see
Tab.~\ref{Tab:Summer:Fits}). 
%\pagebreak[4]


\begin{table}
\caption{\label{Tab:Summer:Fits}Selected best fit parameters for the
  combined, phase averaged \jemx \& \spi spectra and \jemx \& \isgri 
spectra respectively using the \xspec \texttt{CUTOFFPL} model. 
Note that the values are still preliminary due to calibration uncertainties.}

\vspace*{1em}
\renewcommand{\arraystretch}{1.3}
\begin{tabular}{|l|r@{ }l|r@{ }l|}
\multicolumn{1}{l}{Parameter}  & 
\multicolumn{2}{c}{\jemx + \spi} & 
\multicolumn{2}{c}{\jemx + \isgri} \\
\hline
photon index    & 0.51\err{0.10}{0.20} &      & 0.43\err{0.10}{0.16}  & \\
folding energy  & 11.3\err{0.7}{1.3}   & \kev & 12.1\err{0.9}{0.6} & \kev \\
line center     & 53.6\err{3.4}{1.8}   & \kev & 51.9\err{2.4}{2.6} & \kev \\
line $\sigma$   & 7.3\err{1.8}{1.3}    & \kev & 4.0\err{3.5}{3.7}  & \kev \\
line depth      & 0.63\err{0.13}{0.08} &      & 0.38\err{\infty}{0.16}& \\
\redchi with line & 1.19               &      & 1.02 & \\    
\redchi w/o  line & 1.85               &      & 1.17 & \\    
\hline
\end{tabular}
\end{table}

Unfortunately, the current cali\-bration and cross-cali\-bration
uncertainties for the \intg instruments do not allow any firm
statement about the existence or not of a \ca 25\,\kev line in
the spectra. All spectral modeling has been done without including
such a feature but a feature of the strength reported, e.g., by
\citet{Kreykenbohm:2002} would be fully consistent with the data
at the moment.

%\clearpage

\begin{figure*}
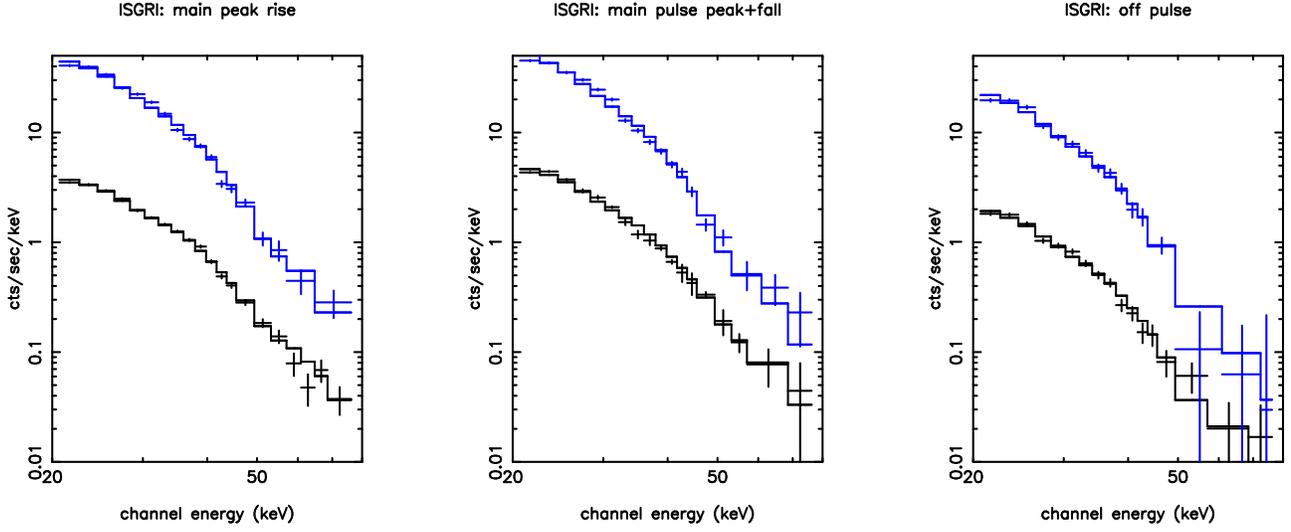

\includegraphics[width=0.28\textwidth]{isgri_rise_compare_cutoffpl.ps}
\hfill 
\includegraphics[width=0.28\textwidth]{isgri_fall_compare_cutoffpl.ps}
\hfill 
\includegraphics[width=0.28\textwidth]{isgri_off1_compare_cutoffpl.ps}

\caption{\label{Fig:Winter_I}Comparison of \isgri spectra before and during the
         first large flare of \vela. From left to right the panels show the spectra
         for the rising flank of the main pulse, the peak and falling flank and
         the following off-pulse region. In each panel the lower spectrum is that
         of the pre-flare while the upper is taken over the flare. 
         The spectral softening is directly visible.}
\end{figure*}

\begin{table*}
\vspace*{-1mm}
\caption{\label{Tab:Winter:ISGRI}Selected best fit parameters for 
 phase resolved \isgri spectra before and during the large
flare of \vela in revolution 137, using the \xspec \texttt{CUTOFFPL} model.
Note that the best fit values are still preliminary.}

\renewcommand{\arraystretch}{1.5}

\begin{tabular}{|l|r@{ }l|r@{ }l||r@{ }l|r@{ }l||r@{ }l|r@{ }l|}
\multicolumn{1}{l}{ } & 
\multicolumn{4}{c}{Main pulse rise} &
\multicolumn{4}{c}{Main pulse peak \& fall} &
\multicolumn{4}{c}{Off pulse } \\[-3pt]
\multicolumn{1}{l}{Parameter} & 
\multicolumn{2}{c}{pre-flare} & \multicolumn{2}{c}{flare} &
\multicolumn{2}{c}{pre-flare} & \multicolumn{2}{c}{flare} &
\multicolumn{2}{c}{pre-flare} & \multicolumn{2}{c}{flare} \\
\hline
norm &
0.13\err{0.01}{0.01} && 2.14\err{0.62}{0.05} &&
0.21\err{0.03}{0.02} && 2.88\err{0.09}{0.07} &&
0.09\err{0.03}{0.01} && 1.43\err{0.25}{0.29} &\\ 
\hline
%photon index &
%0.43 && 0.43 && 0.43 && 0.43 && 0.43 && 0.43 \\
%\hline
folding energy & 
11.2\err{0.5}{0.3} & keV & 9.6\err{0.1}{0.6} & keV &
10.0\err{0.5}{0.3} & keV & 8.9\err{0.1}{0.1} & keV &
 9.6\err{0.4}{0.7} & keV & 8.5\err{0.6}{0.3} & keV \\
\hline
\end{tabular}
\end{table*}

\begin{figure*}
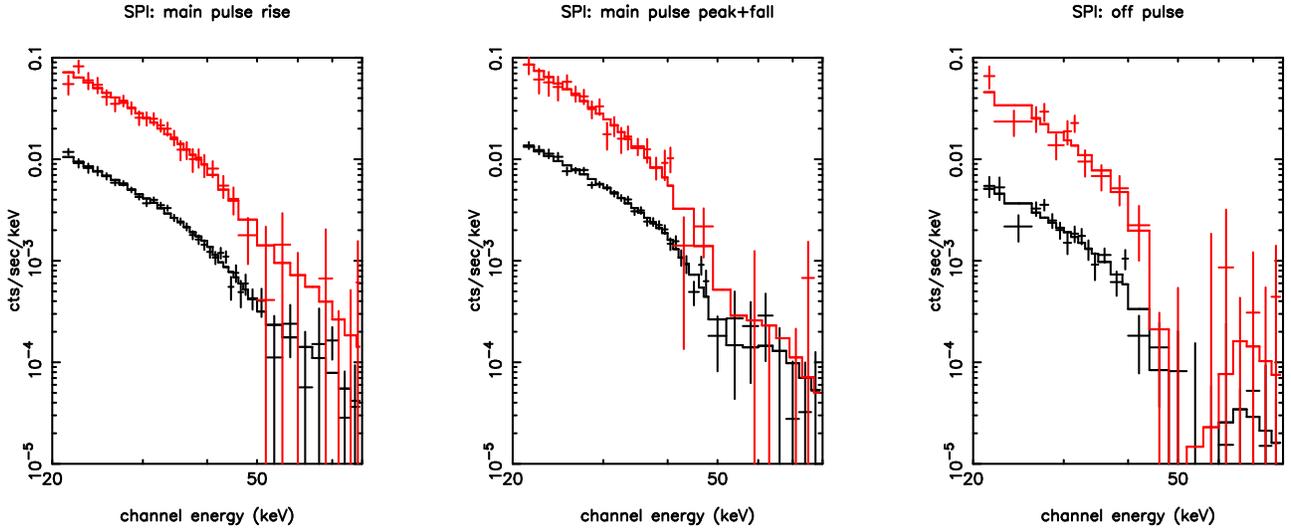

\vspace*{2mm}
\includegraphics[width=0.28\textwidth]{spi_rise_compare_cutoffpl.ps}
\hfill 
\includegraphics[width=0.28\textwidth]{spi_fall_compare_cutoffpl.ps}
\hfill 
\includegraphics[width=0.28\textwidth]{spi_off1_compare_cutoffpl.ps}

\caption{\label{Fig:Winter_S}Comparison of \spi spectra before and during the
         first large flare of \vela. The panels follow the same pattern as in
         \protect{Fig.~\ref{Fig:Winter_I}}. 
         Due to the relatively short integration
         times, the feature at \ca 50\,\kev is not clearly visible.}
\end{figure*}

\begin{table*}
\vspace*{-1mm}
\caption{\label{Tab:Winter:SPI}Selected best fit parameters for 
 phase resolved \spi spectra before and during the large
flare of \vela in revolution 137, using the \xspec \texttt{CUTOFFPL} model.
Note that the best fit values are still preliminary.}

\renewcommand{\arraystretch}{1.5}

\begin{tabular}{|l|r@{ }l|r@{ }l||r@{ }l|r@{ }l||r@{ }l|r@{ }l|}
\multicolumn{1}{l}{ } & 
\multicolumn{4}{c}{Main pulse rise} &
\multicolumn{4}{c}{Main pulse peak \& fall} &
\multicolumn{4}{c}{Off pulse } \\[-3pt]
\multicolumn{1}{l}{Parameter} & 
\multicolumn{2}{c}{pre-flare} & \multicolumn{2}{c}{flare} &
\multicolumn{2}{c}{pre-flare} & \multicolumn{2}{c}{flare} &
\multicolumn{2}{c}{pre-flare} & \multicolumn{2}{c}{flare} \\
\hline
norm &
0.25\err{0.05}{0.04} && 2.17\err{0.81}{0.58} &&
0.32\err{0.07}{0.05} && 4.16\err{2.50}{1.56} &&
0.13\err{0.09}{0.05} && 1.59\err{2.10}{0.90} &\\ 
\hline
%photon index &
%0.43 && 0.43 && 0.43 && 0.43 && 0.43 && 0.43 \\
%\hline
folding energy & 
11.4\err{0.8}{0.8} & keV & 10.2\err{1.1}{1.0} & keV &
11.5\err{0.8}{0.8} & keV & 8.3\err{1.2}{1.0} & keV &
11.0\err{2.5}{1.8} & keV & 9.6\err{3.3}{2.1} & keV \\
\hline
\end{tabular}
\end{table*}

%\clearpage

\subsection{Winter observations}\label{Res:winter}
For the pre-flare and flare data of the winter observations we
are restricted to spectra from the two main instruments as explained
in Section~\ref{Obs}. Fig.~\ref{Fig:Winter_I} and Fig.~\ref{Fig:Winter_S}
show for comparison the phase resolved spectra of before and during 
the first big flare. While the flux increases by a factor of about 10
during the flare, the spectral shape stays more or less the same, except
for a certain softening, which is directly visible in the count spectra.

Using the same model as for the summer observations
the folding energy parameter, which can be seen as a measure of
temperature, decreases by 1--1.5\,\kev during the flare,
see Tab.~\ref{Tab:Winter:ISGRI} and Tab.~\ref{Tab:Winter:SPI}.
For these fits the photon index of the continuum and the line
width and energy were ill determined and fixed to the values 
found for the summer data, see Tab.~\protect{\ref{Tab:Summer:Fits}}.
This had little effect on the continuum parameters.

\section{Discussion}\label{Disc}
While the remaining uncertainties in the calibration of the individual
instruments as well as in the intercalibration do not allow any firm
statement about a possible line feature at 20-25\,\kev we clearly observe
and resolve -- thanks to the high spectral resolution of \spi -- the
well-known line feature above 50\,\kev. The half-width of 
7.2\err{1.5}{1.2}\,\kev
obtained from \spi data is in the typical range of line widths reported from
observations with other instruments and expected from thermal broadening.
It falls inbetween the relatively narrow line solution reported by
\citet{Coburn:2002} and the very broad line width given in
\citet{Orlandini:98}. It should also be kept in mind that this line width
is obtained for a long-term time averaged spectrum, as for shorter
integration times the statistics do not allow to constrain the fluxes in
this energy range anymore.

Based on the theoretical cyclotron line shapes proposed, e.g., by
\citet{ArayaHarding:99}, the broad, flat shape and lack of additional 
structure could be taken as indication that this line is indeed a harmonic
and not the more structured fundamental. Alternatively, this would
indicate observation  at a relatively narrow angle to the magnetic field 
and preferably of a cylindrical emission region.

It is interesting to see that despite a rapid, strong brightening, the
source spectrum appears to remain very similar to that obtained during
more ususal flux levels. With a normal luminosity of 
\ca 4$\times$10\e{36}\,ergs/s the wind-accreting source remains in the 
sub-Eddington regime of accretion even during the flares. Given the well
known pulse to pulse variations \citep{Staubert:80} one can speculate that
the normal mode of accretion for \vela is a relatively sparse and
inhomogenous flow which does not fill the accretion column completely.  In
this picture the spectral softening observed could be caused by additional
comptonisation of X-ray photons by a more densely filled accretion column.

A more detailed analysis of this extensive and higly interesting dataset
than can be done for these proceedings is surely called for in the future.
With further improved flux reconstruction and calibration information not 
only the question of the 20-25\,\kev line can be settled but we will also
arrive to do detailed quantitative studies of the spectral evolution 
in calm and flaring times.

\renewcommand{\refname}{\textbf{References}}
\parskip0pt
\bibsep0pt
\bibliographystyle{aa}
\bibliography{mnenomic,vela,accretion,cyclotron,diverse,instrum,crossref}

\begin{thebibliography}{24}
\expandafter\ifx\csname natexlab\endcsname\relax\def\natexlab#1{#1}\fi
\expandafter\ifx\csname url\endcsname\relax
  \def\url#1{{\tt #1}}\fi
\expandafter\ifx\csname urlprefix\endcsname\relax\def\urlprefix{URL }\fi

\bibitem[{Araya \& Harding(1999)}]{ArayaHarding:99}
Araya R.A., Harding A.K.1999, ApJ, 517, 334

\bibitem[{{Atti{\' e}} et~al.(2003){Atti{\' e}}, {Cordier}, {Gros}
  et~al.}]{Attie:2003}
{Atti{\' e}} D., {Cordier} B., {Gros} M., et~al.2003, A\&A, 411, L71

\bibitem[{{Barziv} et~al.(2001){Barziv}, {Kaper}, {Van Kerkwijk}, {Telting}, \&
  {Van Paradijs}}]{Barziv:2001}
{Barziv} O., {Kaper} L., {Van Kerkwijk} M.H., {Telting} J.H., {Van Paradijs}
  J.2001, A\&A, 377, 925

\bibitem[{{Choi} et~al.(1996){Choi}, {Dotani}, {Day}, \& {Nagase}}]{Choi:96}
{Choi} C.S., {Dotani} T., {Day} C.S.R., {Nagase} F.1996, ApJ, 471, 447

\bibitem[{{Coburn} et~al.(2002){Coburn}, {Heindl}, {Rothschild}
  et~al.}]{Coburn:2002}
{Coburn} W., {Heindl} W.A., {Rothschild} R.E., et~al.2002, ApJ, 580, 394

\bibitem[{Haberl \& White(1990)}]{HaberlWhite:90}
Haberl F., White N.1990, ApJ, 361, 225

\bibitem[{Inoue et~al.(1984)Inoue, Ogawara, Ohashi et~al.}]{Inoue:84}
Inoue H., Ogawara Y., Ohashi T., et~al.1984, PASJ, 36, 709

\bibitem[{Kendziorra et~al.(1990)Kendziorra, Mony, Maisack
  et~al.}]{Kendziorra:90}
Kendziorra E., Mony B., Maisack M., et~al.1990, In: Proc.\ of the 23rd ESLAB
  Symposium on Two Topics in X-ray Astronomy, ESA SP-296, (1)467--471, ESA
  Publications Division

\bibitem[{Kendziorra et~al.(1992)Kendziorra, Mony, Kretschmar
  et~al.}]{Kendziorra:92}
Kendziorra E., Mony B., Kretschmar P., et~al., 1992, In:
  Tanaka Y., Koyama K. (eds.), Frontiers of X-Ray Astronomy, 
  Frontiers Science Series 2, Tokyo, 51--52

\bibitem[{Kretschmar et~al.(1997)Kretschmar, Pan, Kendziorra
  et~al.}]{Kretschmar:97}
Kretschmar P., Pan H.C., Kendziorra E., et~al.1997, A\&A, 325, 623

\bibitem[{Kreykenbohm et~al.(1999)Kreykenbohm, Kretschmar, Wilms
  et~al.}]{Kreykenbohm:99}
Kreykenbohm I., Kretschmar P., Wilms J., et~al.1999, A\&A, 341, 141

\bibitem[{{Kreykenbohm} et~al.(2002){Kreykenbohm}, {Coburn}, {Wilms}
  et~al.}]{Kreykenbohm:2002}
{Kreykenbohm} I., {Coburn} W., {Wilms} J., et~al.2002, A\&A, 395, 129

\bibitem[{{Krivonos} et~al.(2003){Krivonos}, {Produit}, {Kreykenbohm}
  et~al.}]{Krivonos:ATel211}
{Krivonos} R., {Produit} N., {Kreykenbohm} I., et~al.2003, The Astronomer's
  Telegram, 211, 1

\bibitem[{Lapshov et~al.(1992)Lapshov, Sunyaev, Chichkov et~al.}]{Lapshov:92}
Lapshov I.Y., Sunyaev R.A., Chichkov M.A., et~al.1992, Sov.\ Astron.\ Lett.,
  18, 16

\bibitem[{Makishima et~al.(1992)Makishima, Mihara, Nagase, \&
  Murakami}]{Makishima:92}
Makishima K., Mihara T., Nagase F., Murakami T., 1992, In:
  Tanaka Y., Koyama K. (eds.), Frontiers of X-Ray Astronomy, 
  Frontiers Science Series 2, Tokyo,, 23--32

\bibitem[{McClintock et~al.(1976)McClintock, Rappaport, Joss
  et~al.}]{McClintock:76}
McClintock J., Rappaport S., Joss P., et~al.1976, ApJ, 206, L99

\bibitem[{Nagase(1989)}]{Nagase:89}
Nagase F.1989, PASJ, 41, 1

\bibitem[{Nagase et~al.(1986)Nagase, Hayakawa, \& Sato}]{Nagase:86}
Nagase F., Hayakawa S., Sato N.1986, PASJ, 38, 547

\bibitem[{{Orlandini} et~al.(1998){Orlandini}, {Dal Fiume}, {Frontera}
  et~al.}]{Orlandini:98}
{Orlandini} M., {Dal Fiume} D., {Frontera} F., et~al.1998, A\&A, 332, 121

\bibitem[{{Sako} et~al.(1999){Sako}, {Liedahl}, {Kahn}, \& {Paerels}}]{Sako:99}
{Sako} M., {Liedahl} D.A., {Kahn} S.M., {Paerels} F.1999, ApJ, 525, 921

\bibitem[{Staubert et~al.(1980)Staubert, Kendziorra, Pietsch
  et~al.}]{Staubert:80}
Staubert R., Kendziorra E., Pietsch W., et~al.1980, ApJ, 239, 1010

\bibitem[{Staubert et~al.(2004)Staubert, Kreykenbohm, Kretschmar
  et~al.}]{Staubert:IWS5}
Staubert R., Kreykenbohm I., Kretschmar P., et~al.2004, In: The INTEGRAL
  Universe, no. SP-552 in ESA, ESA Publications Division, ESTEC, Noordwijk, The
  Netherlands

\bibitem[{Tanaka(1986)}]{Tanaka:86}
Tanaka Y.1986, In: Mihalas D., Winkler K. (eds.) Radiation Hydrodynamics in
  Stars and Compact Objects, 198--221, Springer-Verlag, Berlin

\end{thebibliography}

\end{document}